\documentclass[twocolumn,showpacs,preprintnumbers,%
amsmath,amssymb,aps,showkeys,floatfix]{revtex4}

\usepackage{graphicx}% Include figure files
\usepackage{dcolumn}% Align table columns on decimal point
\usepackage{bm}% bold math

\begin{document}

%\preprint{DRAFT}

\title{Sound attenuation in an unexplored frequency region:
\\Brillouin Ultraviolet Light Scattering measurements in v-SiO$_2$}

\author{P. Benassi,$^{1}$ S. Caponi,$^{2}$ R. Eramo,$^3$
  A. Fontana,$^{2}$ A. Giugni,$^{1}$  M. Nardone,$^{1}$
  M. Sampoli,$^{4}$ and G. Viliani$^{2}$}

\affiliation{$^1$Dip. Fisica and INFM, Universit\`a  dell'Aquila,
I-67100 L'Aquila, Italy,
  \\$^2$Dip. Fisica and INFM, Universit\`a di Trento, I-38050 Povo
  Trento, Italy
  \\$^3$INFM, Dip. Fisica, and LENS, Universit\`a  di Firenze, I-50019
  Sesto Fiorentino, Firenze, Italy,
  \\$^4$INFM and Dip. di Energetica, Universit\`a di
  Firenze, I-50139, Firenze, Italy
}

\date{\today}

\begin{abstract}
We report Ultraviolet Brillouin light scattering experimental data
on v-SiO$_2$ in an unexplored frequency region, performed with a
newly available spectrometer up to exchanged wavevector $q$ values
of $0.075~nm^{-1}$, as a function of temperature. The measured
attenuation scales on visible data following a $q^2$ behavior and
is temperature-dependent. Such temperature dependence is found in
a good agreement with that measured at lower $q$, suggesting that
its origin is mainly due to a dynamic attenuation mechanism. The
comparison between the present data with those obtained by
Inelastic X-ray Scattering suggests the existence of a cross-over
to a different attenuation regimes.

\end{abstract}

\pacs{78.35.+c, 63.50.+x,62.65.+k}

\keywords{Brillouin Scattering, Disordered Systems, Vibrational States}

\maketitle

\begin{figure}[tbp!]
\includegraphics[width=8cm]{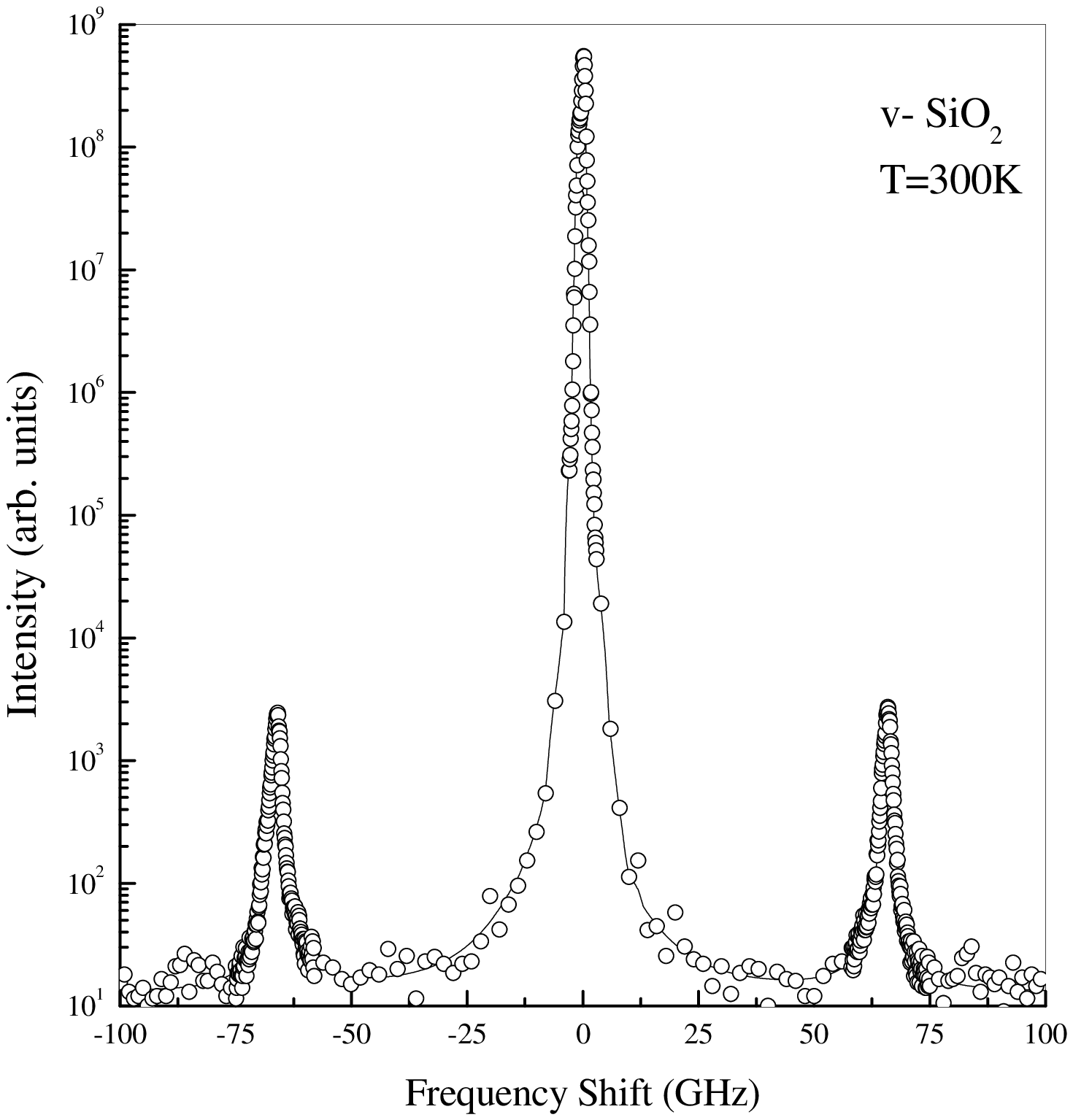}
\caption{Typical Brillouin ultraviolet Light scattering spectrum
on v-SiO$_2$  at $T=300~K$ ($180^\circ$ scattering geometry and
the excitation wavelength was $266~nm$). The experimental data
(circles) are reported together with the best fit (full line) see
text. The excitation wavelength was $266~nm$, with a power
$\approx 100~mW$.}
% The UV beam was focused with a
%$200~mm$-focal lenght lens while the scattering collection angle
%is the full value of the spectrometer; the instrumental resolution
%was set to $0.8~GHz$.
\label{fig1}
\end{figure}

Our present understanding of the sound attenuation mechanisms in
vitreous systems is poor compared to that of crystalline materials
although this topic has attracted the interest of several
researchers from both experimental and theoretical points of view
\cite{Philmag}. In particular, the nature of vibrational dynamics
of disordered systems, as derived from the study of the
"low-frequency" excitations (in the hypersonic range), has been a
highly debated subject in the last years
\cite{Leadbetter,Ruocco,Fabian,Buchenau,Dove}. Vitreous silica
(v-SiO$_2$, or amorphous quartz) has been extensively studied as
prototype of a strong glass. Nevertheless, the experimental
results reported in literature had conflicting and controversial
interpretations \cite{Benassi1,Foret,Rat,Elliott2,Pilla1}. To
achieve an overall characterization of this particular system is
crucial to understand the vibrational dynamics of the vitreous
state. It is well known that a plane wave excitation can propagate
freely in a disordered structure only when the wavelength is much
greater than the scale spanned by microscopic inhomogeneities. As
the wavelength shortens the wave is attenuated, distorted, and
scattered with increasing magnitude. As a matter of fact, the
waves can even become over-damped when the wavelength approaches
the interparticle spacing and the excitations become more directly
affected by the microscopic disorder characteristic of the glass
structure. The question about the nature of excitations far from
the long wavelength limit, is unlikely to have a single answer
\cite{Foret1,Caponi,Taraskin1}.  Indeed different mechanisms have
been suggested in order to explain the attenuation of an acoustic
plane wave excitation namely: attenuation induced by topological
disorder, thermally activated processes, anharmonic effects,
two-level systems (TLS) processes \cite{Hunklinger}. In general,
anharmonic effects are particularly relevant at relatively high
temperatures while the importance of the two-level systems is
confined at very low temperatures ($T<5\div 10~K$). In the
intermediate range $10~K<T\ll T_g$ the TLS effects are not
important and disordered induced mechanisms, thermally activated
processes and anharmonic effects play the most important role in
acoustic attenuation. Anyway, sound propagates through glasses and
longitudinal and transverse phonons are found to be still well
defined at relatively large wave-vectors (up to $5~nm^{-1}$) with
a linear relationship between frequency ($\omega$) and exchanged
wavevector ($q$) \cite{Ruzicka,Pilla2}. Beyond this limiting $q$
value, the effect of the topological disorder of the glassy
systems causes (i) a mixing of the polarisation of the acoustic
modes, which is observed for different glass forming systems both
in experimental \cite{acquasette,glicerolo,Ruzicka} and in
molecular dynamic  simulations (MDS) data
\cite{acquasampoli,Pilla2} and (ii) a possible presence of
positive dispersion in the longitudinal acoustic branch
\cite{Ruzicka}, as theoretically predicted \cite{Goetze} and found
in MDS \cite{Pilla2}. Since below such $q$ values, v-SiO$_2$
exhibits a substantially constant sound velocity, so the
dependence of the sound attenuation on $\omega$ can be safely
investigated by changing the wavevector $q$. \newline
 The acoustic attenuation can
be measured by the full width half maximum, $\Gamma$, of the
Brillouin doublet characterizing the dynamical structure factor
$S(q,\omega )$. In the hypersonic region (i.e. in the GHz region
investigated by the Brillouin Light Scattering technique, BLS),
$\Gamma$ exhibits a strong temperature dependence and suggests
that the attenuation at these frequencies has to be mostly
ascribed to dynamical processes \cite{Vacher,Vacher2}. On the
contrary, in the mesoscopic range (i.e. in the THz region
investigated by the Inelastic X-ray Scattering technique, IXS)
$\Gamma$ has a negligible temperature dependence, supporting a
non-dynamic origin of acoustic attenuation in this frequency
regime \cite{Ruocco}.
 In vitreous silica, the prototype of
strong glasses \cite{Angell}, two main attenuation mechanisms have
been hypothesized in the GHz and THz regions. In the first
scenario \cite{Vacher, Foret2, Ruffle} the attenuation is
characterized by a crossover from a mechanism dominated by
dynamical relaxation processes with a frequency dependence of
$\Gamma \propto \omega ^2 \propto q^2$ in the GHz range, to a
mechanism dominated by strong phonon scattering, due to presence
of topological disorder in the THz region. It is hypotized that
the last process should exhibit a $\Gamma \propto \omega ^4
\propto q^4$ frequency dependence (much like the Rayleigh
light-scattering regime by independent particles). The $\Gamma
\propto \omega ^4$ behaviour is able to explain the plateau in the
conductivity measurements. The second scenario \cite{Ruocco} also
suggests the existence of a crossover between a low frequency
(dynamical origin) attenuation mechanism and a high frequency one
(nearly temperature-independent) due to topological disorder.
However a $\Gamma \propto \omega ^2 \propto q^2$ dependence is
hypothesized for both mechanisms
\cite{Sette,Benassi1,Pilla2,Ruocco,acquasette}. Such a $\Gamma
\propto \omega ^2 \propto q^2$ dependence valid up to high q, has
also been predicted by molecular dynamic simulations performed on
realistic v-SiO$_2$ \cite{Dellanna}, on harmonic glasses
\cite{Ruocco2} as well as on hard spheres systems \cite{Goetze}
and disordered linear chains \cite{Montagna}; it is also supported
by recent theoretical free-energy landscape studies
\cite{Parisi1,Parisi2}.
\newline BLS and IXS technique do not cover the entire frequency (and
$q$) range from GHz to THz and investigations within the frequency
gap which separates these techniques could be useful to
discriminate between the different hypotheses. As a matter of
fact, between the GHz and THz region, the sound attenuation has
already been measured by the picosecond optical technique (POT)
\cite{Zhu} and no dependence on temperature of $\Gamma$ has been
found in the range $80\div 300~K$. Unfortunately, these difficult
measurements are affected by large systematic errors and random
uncertainties. In the region where there is an overlapping with
the BLS attenuation data, the POT values of $\Gamma$ are always
more than a factor $2$ larger. As a consequence, even if the POT
data can give a qualitatively insight of the q behaviour, they are
of little help in understanding the different behaviour of the
temperature dependence of $\Gamma$ found by BLS and IXS
techniques. In order to shed some light on the underlying
mechanisms an accurate investigation of the attenuation in the
intermediate $q$-region. This is now possible thanks to the
development of a new spectroscopic apparatus operating in the
ultraviolet region \cite{Benassi2}. Here we report the
investigation of ultraviolet Brillouin light scattering (BUVS) on
v-SiO$_2$ (spectrosil) performed at different temperatures. We
find that, in the BUVS regime, the measured line-width $\Gamma$
appears to be temperature dependent with a $q^2$ behaviour. Our
findings are in agreement with the existence of two main
attenuation mechanisms, with a relative weight which is dependent
on the phonon wavelength. They are also compatible with the
existence of a crossover between a prevalent dynamic and a
prevalent static sound attenuation mechanism and give some hints
for a possible explanation of  the $\Gamma$ excess found by
extrapolating the $q^2$ behavior of IXS data down to BLS. \newline
 The experimental apparatus used for the
measurements with ultraviolet excitation consists of a new-built
spectrometer with high resolution, contrast, and luminosity,
called HIRESUV \cite{Benassi2}. The instrument is based on a
double grating 4 meter focal length monochromator specifically
designed for Brillouin spectroscopy with both visible ($532~nm$)
and ultraviolet ($266~nm$) excitation. The high luminosity is
achieved with the two large echelle grating (400x208 mm with
$31.6~grooves/mm$).  The size of the gratings and mirrors yield an
instrumental $F-$number of $1:40$ in the vertical plane (where
dispersion occours) and of $1:20$ in the horizontal plane. In the
UV range, HIRESUV works at the $230$-th order and reaches a
resolution of about $0.6~GHz$. To overcome air turbulence effects
and small changes of refractive index and further to warrant a
good thermal conductivity, the entire apparatus, except the sample
compartment, is inside a vacuum chamber filled with helium at
atmospheric pressure. All the optical items (mirror, slits, and
light collecting lenses) are positioned and aligned by means of
computer controlled microstep movements. The photocounts of a low
noise phototube are recorded at fixed wavelength for subsequent
elaboration. The UV radiation is generated by a commercial system
based on the second-harmonic generation of  a visible laser
source. Other details of HIRESUV will be published elsewhere
\cite{Benassi2}. A typical spectrum obtained using HIRESUV at room
temperature is reported in fig.~\ref{fig1} (the measured
resolution in this case was set to about $0.8~GHz$) where the high
contrast and high resolution  reached by the experimental
apparatus can be appreciated. The BUVS signal (open circles),
proportional to the dynamic structure factor, $S(q, \omega )$,
convoluted with the instrumental resolution function, $R(\omega
)$, is reported in a log scale together with a fit function (full
line). To get quantitative information on characteristic frequency
and on the attenuation $\Gamma$ of the excitations, the data have
been fitted by the convolution of the experimentally determined
$R(\omega )$ with the sum of an elastic and an inelastic
contribution; the former, is represented by a $\delta$-function
while the latter has been described by a DHO model.

\begin{figure}[tbp!]
\includegraphics[width=8cm]{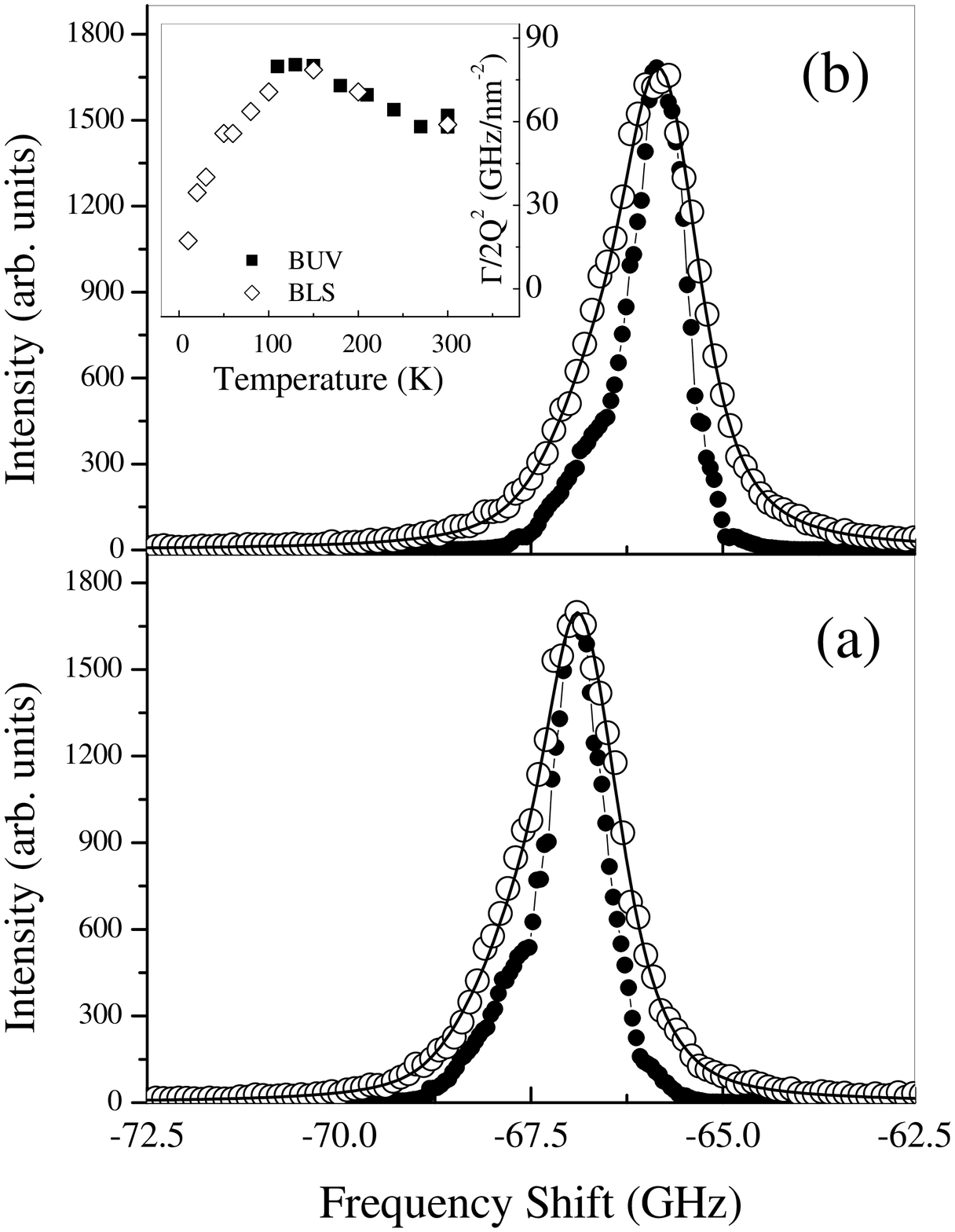}
\caption{ Anti-Stokes part of BUVS spectra (circles) at two
significant temperatures: a) at $270~K$, b) at $150~K$. The
resolutions (full lines and dots) and the best fit (full lines)
are also reported. In the inset is shown the temperature behaviour
of the $\Gamma/2q^2$ for the BLS (open diamonds)and BUVS (full
squares) data. } \label{fig2}
\end{figure}

\begin{figure}[tbp!]
\centering
\includegraphics[width=5cm] {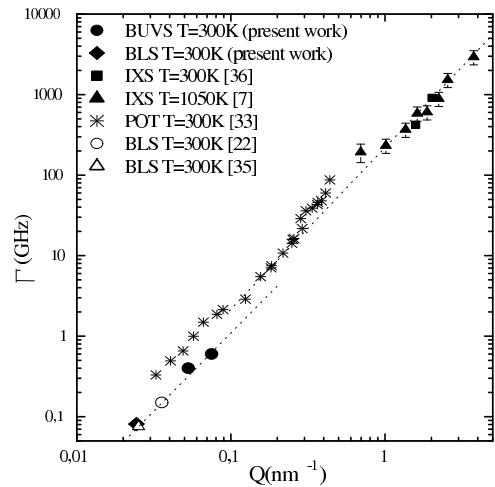}
\caption{ Log-log plot of Brillouin widths as a function of
exchanged wavevector $q$ obtained at room temperature by BLS (full
diamond, open triangle \cite{Bucaro},  open circle \cite{Vacher}),
BUVS (full circle) in $90^\circ$ and $180^\circ$ scattering
configurations, and IXS  (triangles $T=1050~K$, squares
$T=300~K$). Asterisks represent the results obtained by POT
technique \cite{Zhu}. The dashed lines represent the $q^2$ law for
the respective data.} \label{fig3}
\end{figure}

As an example of the temperature behaviour of the Brillouin
peak, the Anti-Stokes parts of the spectrum for two selected
temperatures (open circles) together with their resolution curves
(dots connected by lines) and the best fit (full lines) are
reported in fig.~\ref{fig2}. 
Directly from the rough BUVS data one
can visualize the temperature changes of the width, indicating
that, in this frequency range, at least part of the sound
attenuation has a dynamical origin. In the inset of
fig.~\ref{fig1}, the HWHM of the Brillouin peak, once scaled by
$q^2$, as a function of temperature is reported; the temperature
behaviour seems to be in good agreement with the one obtained by
BLS measurements \cite{Vacher}. The values of $\Gamma (q)$
measured at room temperature resulting from the fits of BUVS
(dots) and BLS (full diamond, circle \cite{Vacher} and open
triangle \cite{Bucaro}) data are reported in fig.~\ref{fig3}. 
The
spectra collected at $90^\circ$-scattering angle were measured
with an horizontal collection angle of $\delta \phi _H=1.7~mrad$.
In this configuration the exchange wavevector dispersion arising
from the finite acceptance angle, $\delta q/q=\delta \phi _H/2$
gives a negligible contribute to the the mode linewidth. The
dashed lines represents the extracted $q^2$ behaviour. In the same
figure we have reported the POT data (asterisks \cite{Zhu}) and
IXS data  at $T=1050~K$ (full triangles \cite{Benassi1}) and at
$T=300~K$ (full squares \cite{Mascio}). The new BUVS data are
along the same BLS trend while BLS and BUVS data are always more
than a factor $2$ below the POT data. In our opinion, this
confirms that the POT data can only give a qualitative
information. At high $q$, inside the experimental error, $\Gamma
(q)$ obtained by IXS do not show any noticeable temperature
dependence in the $300\div 1500~K$ investigated range
\cite{Ruocco} and the measured values are in good agreement with
the ones calculated by MDS for the extreme $T=0$ case
\cite{Dellanna}. The $\Gamma (q)$ is well fitted to a $q^2$ law in
the $1\div 5~nm^{-1}$ range confirming, again, the general trend
found in the simulations \cite{Dellanna}. Within the experimental
uncertainties, the temperature independence of $\Gamma$ indicates
that most of the attenuation has a non-dynamical origin and is
ascribable to the presence of topological disorder \cite{Ruocco}.
It is worth to notice that a $q^2$ law is not able to reproduce
accurately both high-q IXS and low-q BLS and BUVS data at room
temperature. Therefore $\Gamma (q)$ must have a more complex
behaviour than a $q^2$ dependence: between ~$1~nm^{-1}$ down to
~$0.1~nm^{-1}$ it should decrease more steeply in order to join
the $q^2$ behaviour at low-q values. Temptatively we will write
the attenuation $\Gamma$ as a sum of two contributions: $\Gamma
(q,T)=\Gamma _s(q)+\Gamma _d(q,T)$, where $\Gamma _s(q)$ is the
structural part dependent on the topological disorder at the
nanometer scale while $\Gamma _d(q,T)$ is due to dynamical
temperature dependent processes effective at the micrometer scale.
The $\Gamma _s(q)$ part is independent on temperature since the
topological structure at molecular level is insensitive to
temperature changes, as confirmed by the measurements of the
static structure factor in many glasses. It is the dominant
contribution at higher $q$ values. On the contrary, when the
wavelength is much larger (in BLS range) and the system is viewed
as a continuum, the disorder produced by dynamical processes
becomes the major responsible for the attenuation and the
contribution related to the structural disorder, $\Gamma _s(q)$,
will be lower than the $\Gamma$ minimum value reached around $5~K$
(about two order of magnitude smaller than the $300~K$ value)
\cite{Vacher}. The interplay between the two mechanisms, in the
intermediate frequency region, results in a cross-over from a
temperature dependent to a temperature independent sound
attenuation, depending on which one of the to processes is the
dominating  mechanisms. The BUVS measurements, performed as a
function of temperature, indicate a dynamic contribution up to
exchange wavevector values of $q\approx 0.075~nm^{-1}$. In
conclusion in this work, we have reported the measure of the
Brillouin spectra performed employing UV excitation on v-SiO$_2$,
using the new HIRESUV spectrometer. The experimental data obtained
in the, up to now, unexplored frequency region give indications
that in vitreous silica the dominant sound attenuation mechanism
has a distinct origin in the different q regions spanned by the
IXS or BUVS/BLS techniques. Notwithstanding the complex behaviour
of $\Gamma (q,T)$, which appear to be far from a simple square
law, the data do not support the first hypothesis on the origin of
sound attenuation \cite{Vacher,
 Foret2, Ruffle}. Indeed, in the high frequency or
exchange wavevector range, the attenuation is never found to be
growing with a fourth power law and moreover, even if in the
unexplored q range (between $0.1~nm^{-1}$ and $1~nm^{-1}$) a
steeper than $q^2$ dependence has to be expected, at high $q$
values the $q^2$ behaviour is again recovered. Concerning the
second hypothesis, other measurements in the unexplored frequency
region are necessary to verify the predicted existence of the
cross-over region. Moreover we want to stress that, in the
framework of second hypothesis, the excited sound plane wave
inside the glass in the THz range is far from a plane wave; on the
contrary it is formed by a large spectral combination of
eigenvectors, and the loss of phase relationship between them is
the major cause of the plane wave attenuation. This attenuation is
not effective for the thermal conductivity which is related to the
decay of a given eigenvector amplitude  and not to a phase change
between different eigenvector components. This could explain
qualitatively why the thermal conductivity behaves so differently
from sound attenuation and further earns high marks for the second
hypothesis.

\end{document}